\documentclass{aa}
\usepackage{graphicx}
\usepackage{txfonts}
\begin{document}
%
   \title{Evolution of magnetic fields in galaxies and future observational tests with the Square Kilometre Array}


   \author{Tigran G.\ Arshakian\inst{1}, Rainer\ Beck\inst{1},
   Marita \ Krause\inst{1}, \and Dmitry\ Sokoloff\inst{2}}

   \offprints{T.G. Arshakian}

   \institute{Max-Planck-Institut f\"ur Radioastronomie, Bonn, Germany\\
              \email{[tarshakian;rbeck;mkrause]@mpifr-bonn.mpg.de}
         \and Department of Physics, Moscow State University, Russia\\
             \email{sokoloff@dds.srcc.msu.su}
             }

   \date{Received September 32, 2099; accepted September 33, 2099}


  \abstract
   {}
   {We investigate the cosmological evolution of large- and small-scale
magnetic fields in galaxies in the light of present models of formation and evolution of galaxies.}
   {We use the dynamo theory to derive the timescales of amplification
and ordering of magnetic fields in disk and puffy galaxies.
Turbulence in protogalactic halos generated by thermal virialization
can drive an efficient turbulent dynamo. Results from simulations of
hierarchical structure formation cosmology provide a tool to develop
an evolutionary model of regular magnetic fields coupled with galaxy
formation and evolution.}
   {The turbulent (small-scale) dynamo was able to amplify a weak seed
magnetic field in halos of protogalaxies to a few $\mu$G strength
within a few $10^8$ yr. This turbulent field served as a seed to
the mean-field (large-scale) dynamo. Galaxies similar to the Milky
Way formed their disks at $z\approx10$ and regular fields of $\mu$G
strength and a few kpc coherence length were generated within 2~Gyr
(at $z\approx3$), but field-ordering on the coherence scale of
the galaxy size required an additional 6~Gyr (at $z\approx0.5$). Giant
galaxies formed their disks at $z\approx10$, allowing more
efficient dynamo generation of strong regular fields (with kpc
coherence length) already at $z\approx4$. However, the age of the
Universe is short for fully coherent fields in giant galaxies
larger than 15~kpc to have been achieved. Dwarf galaxies should have hosted fully coherent fields at $z\approx1$. }
   {This evolutionary scenario can be tested by measurements of
polarized synchrotron emission and Faraday rotation with the planned
Square Kilometre Array (SKA). We predict an anticorrelation between
galaxy size and ratio between ordering scale and galaxy size. Weak regular fields (small Faraday
rotation) in galaxies at $z\la3$ are signatures of major mergers.
Undisturbed dwarf galaxies should host fully coherent fields, giving
rise to strong Faraday rotation signals. 
}


\titlerunning{Evolution of magnetic fields in galaxies}
\authorrunning{T. Arshakian et al.}

\maketitle
%

\section{Introduction}

Polarized synchrotron emission and Faraday rotation inferred the presence of regular
large-scale magnetic fields with spiral patterns in the disks of
nearby spiral galaxies (Beck 2005), which were successfully
reproduced by mean-field dynamo theory (Beck et al. 1996; R\"udiger
\& Hollerbach 2004; Shukurov 2005). It is therefore natural to
apply dynamo theory also in predicting the generation of magnetic
fields in young galaxies at high redshifts. However, there is a lack
of observational information with which to test models. The available data
(Kronberg et al. 2008; Bernet et al. 2008; Wolfe et al. 2008;
see also the review by Kronberg 1994), although limited,
imply that the environments of galaxies are significantly
magnetized at high redshifts ($z\la 3$), with regular magnetic-field
strengths comparable to or higher than those at the current
epoch. Another indication of strong magnetic fields in young
galaxies is the tight radio -- far-infrared correlation, which is
valid in galaxies at least to distances of $z\simeq3$ (Seymour et
al. 2008). To explain the almost linear correlation, the total
magnetic-field strength must be related to the star-formation rate
(Lisenfeld et al. 1996, Niklas \& Beck 1997). Since observations at
high redshifts are biased towards starburst galaxies, strong
magnetic fields presumably do exist in distant objects.

We now have sufficient evidence that strong magnetic fields were
present in the early Universe and that synchrotron emission
should be detected with future radio telescopes such as the Square
Kilometre Array (SKA). The SKA will spectacularly increase the
sensitivity and angular resolution of radio observations and allow us
to observe an enormous number of distant galaxies at similar resolution to that achievable for nearby galaxies today (van der Hulst et al. 2004).

The dynamo was introduced in to astrophysics as a mechanism for
transforming the kinetic energy of the motions of electrically
conductive media into magnetic energy. This concept describes the
amplification of small-scale as well as large-scale, regular fields.
For practical, mathematical reasons, the dynamo equation was
separated into a small-scale and a large-scale part, two aspects of
the same mechanism with widely different timescales, but often
considered as two independent dynamo models.

The amplification of large-scale magnetic fields and the general
growth of magnetic energy are not the only merits of galactic
dynamos. For the evolution of galactic magnetic fields, the
transformation of a turbulent into a regular field is of similar
value. It is the large-scale regular field whose observation is most
spectacular and whose generation has attracted most attention in
astrophysics. As far as we know today, the generation of regular
magnetic fields is a delicate process that requires large-scale
rotation of the body under consideration. This rotation
produces the $\alpha$-effect, which amplifies the magnetic field, and
differential rotation ensures that this amplification is effective. This defines
the classical ``mean-field'' galactic dynamo as discussed
below, taking into account recent results from numerical simulations
of galaxy formation.

Developing a consistent dynamo model to describe the entire lifetime of a galaxy
requires information about the basic physical parameters controlling
dynamo action and their evolution. Regular large-scale magnetic
fields can be generated and amplified by the mean-field galactic
dynamo in high-redshift galaxies, provided that a gaseous, rotating
disk already exists. The formation of disk galaxies and the epoch
of this formation are fundamental problems in astronomy. High
resolution numerical simulations of disk formation in galaxies
demonstrated that a dynamical disk could be formed at redshifts $z\sim
5-6$ and even higher (Governato et al. 2007; Mayer et al. 2008). A
more robust understanding of the history of magnetism in young galaxies
may help to solve fundamental cosmological questions about the
formation and evolution of galaxies (Gaensler et al. 2004).

In the absence of large-scale rotation, turbulent motions can amplify
magnetic fields, which are, in general, irregular, small-scale fields.
This process is therefore known as the {\it turbulent or
small-scale dynamo} (e.g. Zeldovich et al. 1990). The growth in
magnetic fluctuations is clearly visible in many numerical MHD
simulations of the interstellar medium (Meneguzzi et al. 1981;
Hanasz et al. 2004; de Avillez \& Breitschwerdt 2005; Kowal et al.
2006; Iskakov et al. 2007; Gressel et al. 2008; Wang \& Abel 2008).
We note that all numerical simulations of the generation of regular
magnetic fields in disk galaxies begin at the disk formation epoch;
we assume that seed fields of strength between a few $\mu$G and
$10^{-3}$\,$\mu$G exists already in the disk and do not consider
the field amplification up to this level.

High-resolution simulations of protogalactic clouds
demonstrated that significant turbulence can be generated prior to
disk formation during the thermal virialization of the halo (Wise \&
Abel 2007). Strong turbulence at these early epochs may drive a
turbulent dynamo, which can effectively amplify the halo magnetic fields.
The amplified magnetic fields can form the basis of a seed
field for the generation of regular fields in the disk of a newly
formed galaxy.

The first model of dynamo action in young galaxies by Beck et al.
(1994) proposed a two-stage dynamo, a turbulent dynamo to amplify a
weak, seed field in a disk within less than a Gyr and a large-scale
dynamo to amplify the regular field within a few Gyr. However, some
of the assumptions, e.g. the epoch of disk formation, the large disk
thickness and the constant radius, are in conflict with the modern
scenario of galaxy formation.

Dynamo theory has been well developed as a branch of MHD and,
as a first step, we exploit mainly conventional approaches to
galactic dynamos (see e.g. Ruzmaikin et al. (1988), Beck et al. (1996) and
R\"udiger \& Hollerbach (2004)) rather than more recent findings in dynamo
theory (see e.g. Tobias \& Cattaneo 2008). In this paper, we develop
a simplified model for the evolution of magnetic fields in both
protogalactic halos and galaxies based on the recent numerical developments
in the study of the formation and evolution of galaxies during the epoch of hierarchical
structure formation. We use the known timescales of the turbulent
dynamo and mean-field dynamo in puffy and disk galaxies to explore
the earliest generation and evolution of regular magnetic fields in
halos and to describe the main phases of magnetic-field evolution in
dwarf and disk galaxies, as well as the influence of merging events
on the magnetic evolution. In Sect.~\ref{sec:mf_evolution}, we derive
the timescales of field amplification and field ordering in disk and
spherical galaxies by the mean-field dynamo. The timescale of the
turbulent dynamo is discussed in Sect.~\ref{sec:ts_td}, and the
origin of seed magnetic fields in Sect.~\ref{sec:osmf}. In
Sect.~\ref{sec:edg}, we summarize developments in the
formation and evolution of disk and dwarf galaxies, and in
Sect.~\ref{sec:gmfdg}, we develop an evolutionary model of regular
magnetic fields in galaxies. Perspectives on detecting polarized radio
emission from nearby and from distant galaxies are discussed in
Sect.~\ref{sec:discussion}, and conclusions are drawn in
Sect.~\ref{sec:conclusions}.

Throughout the paper, a flat cosmology model
($\Omega_{\Lambda}+\Omega_{m}=1$) is used with $\Omega_{m}=0.3$ and
$H_0=70$ km\,s$^{-1}$\,Mpc$^{-1}$.

\section{Magnetic-field generation in galaxies by the mean-field dynamo}
\label{sec:mf_evolution}

In this Section, we refer to thin-disk galaxies that rotate
differentially. Theoretical analysis of differentially rotating
disks demonstrated that a disk is only locally stable to axisymmetric
perturbations if the Toomre's stability criterion is fulfilled
(Julian \& Toomre 1966), i.e. only cool disks in which both the gas and
stars have low velocity dispersion can survive for a long time (Binney
\& Tremaine 1987). Hence, the velocity dispersion in disk galaxies
at higher redshifts is probably similar to that in nearby galaxies.
In the following, we adopt the usual values of $v=10$ km s$^{-1}$ and
a scale of turbulence of $l=100$ pc. Turbulence in galaxy disks is
assumed to be driven by supernovae (Ruzmaikin et al. 1988; Korpi et
al. 1999; de Avillez \& Breitschwerdt 2005; Gressel et al. 2008).
The other basic parameters of the galactic mean-field dynamo are the
angular velocity of the galactic rotation $\Omega$, the disk
thickness $h$, the disk radius $R$, and the gas density $\rho$.

\subsection{Timescales of regular magnetic fields in disk galaxies}
\label{subsec:dynamo_thin}

The conventional theory of the galactic mean-field dynamo in disk
galaxies allows the generation of a regular galactic magnetic field
as a result of joint action of differential rotation $\Omega$ and
helical turbulent motions of interstellar gas. The latter is
responsible for the so-called $\alpha$-effect ($\alpha$). According
to Ruzmaikin et al. (1988), the intensity of both generators can be
characterized by two dimensionless numbers, $R_\omega$ and
$R_\alpha$:

\begin{equation}
R_\omega = {1 \over \beta} {{\left(h \over 500
\,\mathrm{pc}\right)^2 \left( \Omega \over 20
\,\mathrm{km}\,\mathrm{s}^{-1}\,\mathrm{kpc}^{-1}  \right)}}
\label{Rom}
\end{equation}
and
\begin{equation}
R_\alpha = {\alpha \over \beta}{\left(h \over 500
\,\mathrm{pc}\right)}, \label{Ralpha}
\end{equation}
where $\Omega$ and $h$ are the angular velocity of the galaxy and
the half-thickness of a galactic disk (normalized to the galactic
angular velocity and half-thickness, respectively) and $\beta$ is
the turbulent diffusivity. A flat rotation curve ($r
\partial \Omega /\partial r \sim \Omega$) was assumed in deriving
the above relations. The parameters $\alpha$ and $\beta$ cannot be measured
directly from observations, and our aim is to reduce them to
observable parameters. According to simple mixing length theory the
turbulent diffusivity is

\begin{equation}
\beta = {{lv} \over 3} \label{mislbet}
\end{equation}
where $l$ is the basic length scale of turbulence, $v$ is the turbulent
velocity, and the numerical factor characterizes the dimensions of the
space. Typical numbers, as found from observations of nearby
galaxies, are $l=100$ pc and $v=10$ km s$^{-1}$ (e.g. Ruzmaikin et
al. 1988). The $\alpha$-effect is determined by the Coriolis force
and density gradient (Krause \& R\"adler 1983):

\begin{equation}
\alpha = {{\Omega l^2} \over h}. \label{Krause}
\end{equation}

The joint action of both generators can be described by a
so-called dynamo number:

\begin{equation}
|D_{\mathrm d}| = R_{\omega}R_{\alpha} \simeq 9 \left({{h \Omega}
\over v}\right)^2. \label{dynnum}
\end{equation}
The absolute value is taken because $D_{\mathrm d}$ is usually
considered to be negative. The dynamo growth rate ($\Gamma$) is
given by
\begin{equation}
\Gamma = |D_{\mathrm d}|^{1/2} {\beta \over {h^2}}, \label{growth}
\end{equation}
where $\beta / h^2$ is the time of turbulent diffusion. The
\emph{dynamo timescale} for amplification of the regular field
is $t^* = \Gamma^{-1}$ (e-folding time), and we derive

\begin{equation}
  t^* = {h \over {\Omega l}}
  \label{eq:ts_dd}
\end{equation}
from Eqs.~(\ref{mislbet}), (\ref{dynnum}) and (\ref{growth}).
This estimate is the result of combining standard estimates
from galactic dynamo theory, which has not been presented before.

During its lifetime $T$, a galaxy completes $N$ rotations. Assuming
$\Omega \approx
20\,\mathrm{km}\,\mathrm{s}^{-1}\,\mathrm{kpc}^{-1}$, $T=10^{10}$
years and $h/l \approx 5$, we derive $N\approx 30$ and $t^* \approx
5T/(2 \pi N) \approx 2.5 \times 10^{8}$ yr. Hence, the conventional
galactic mean-field dynamo can amplify the regular magnetic field by
a factor of at most $10^{17}$, while in practice it can be lower (see
Sect.~4).

The galactic dynamo is a threshold phenomenon. The regular magnetic
field is only maintained (or amplified) if the dynamo number is
larger than a critical value $|D_{\mathrm d}| > |D_{\mathrm
{cr,\,d}}|$. Ruzmaikin et al. (1998) derived $|D_{\mathrm {cr,\,d}}| \simeq 7$ from the numerical simulations of galactic
dynamo models, which they determined to be the minimal value required for the
dynamo effect to overcome turbulent diffusion. If $|D_{\mathrm d}| <
|D_{\rm {cr,\,d}}|$, the regular magnetic-field decays. The typical
value of dynamo number for the Milky Way (called MW hereafter) is
$|D_{\mathrm d}| \simeq 9$ (note that $h \Omega /v\approx 1$ for the
MW), indicating that the dynamo mechanism acts in the MW. If $h
\Omega/v \ll 1$, a disk galaxy cannot amplify or maintain regular
magnetic fields by means of the mean-field dynamo mechanism.

The expression for $t^*$ (Eq.~\ref{eq:ts_dd}) can be written in
terms of the magnetic-field strength. The time required to amplify
the large-scale magnetic field from $B_0$ to $B_*$ is given by

\begin{equation}
\hat t = t_* \ln (B_*/B_0); \label{hatt}
\end{equation}
If $\hat t \ge T$, the dynamo is in its kinematic regime until the
present time. Hence, the minimum seed field required for dynamo
action is

\begin{equation}
B_0 = B_* \exp (-T/t^*). \label{seedest}
\end{equation}

We note that $t^*$ is the e-folding timescale for amplification of the
magnetic field. The resulting magnetic field is spatially ordered on
a coherence scale $l_{\rm c}$ of a few kiloparsecs (kpc), depending on the
disk half-thickness. Full ordering of the field on the coherence
scale of the radius $R$ ($l_{\rm c}=R$) of a galaxy takes a far longer time.
According to Moss et al. (1998), this \emph{ordering timescale}
$\tilde t$ is

\begin{equation}
\tilde t = {l_{\rm c} \over {\sqrt{\Gamma \beta}}} = {{h^2} \over \beta} {l_{\rm c}
\over h} \, |D_{\mathrm d}|^{-1/4}. \label{eq:ts_o}
\end{equation}

Magnetic-field ordering on the scale of $l_{\rm c}$ is not simply an
instability that can be adequately described by an e-folding time
but a more delicate process called ``front propagation'' (the
so-called Kolmogorov-Petrovsky-Piskunov effect, Kolmogorov et al.
1937; for review in the context of dynamo theory, see also Zeldovich
et al. (1990)). For a crude estimate, one can use $\tilde t$ from
Eq.~(\ref{eq:ts_o}) in the sense that, for times $t < \tilde t$, the
ordering remains in progress, while for $t
> \tilde t$, the ordering process is already finalized.

Equation~(\ref{eq:ts_o}) can be presented in terms of
diffusion timescales in the vertical ($T_{h}$) and radial ($T_{r}$)
directions. The timescale for full ordering $\tilde t(l_{\rm c}=R)$ is far longer than
the amplification timescale $t^*$ (see factor $R/h$ in
Eq.~\ref{eq:ts_o}) as well as the diffusion timescale in the vertical
direction:

\begin{equation}
  T_h = {{h^2} \over \beta} \simeq 7 \, 10^8 {\,\rm yr},
\label{perp}
\end{equation}
and $\tilde t \ll T_{r}$,

\begin{equation}
  T_r = {{R^2} \over \beta} \simeq 400 \, T_h \gg T.
  \label{par}
\end{equation}
The ordering timescale is governed by the geometrical mean of both timescales

\begin{equation}
  \tilde t (l_{\rm c}=R)= \sqrt{T_hT_R} \,\, |D_{\mathrm d}|^{-1/4} \approx \frac{R}{l}
  {\left( h \over v\Omega \right)}^{1\over2}.
\label{geommean}
\end{equation}

For MW-type galaxies, $\tilde t$ is only a few times shorter than the
galaxy lifetime $T$, indicating that transient field configurations,
substantially different from the leading eigenmode of the galactic
dynamo, can survive in some galaxies until now. This may explain
e.g. the large-scale field reversals observed in the MW.

After the ordering timescale $\tilde t$, a steady-state magnetic
field is maintained. The configuration of this field is expected to
be similar to the leading eigenmode of the galactic dynamo,
saturated by nonlinear effects (see Sect.~\ref{subsec:emfs}).

\subsection{Equilibrium magnetic-field strength in disk galaxies}
\label{subsec:emfs}

The initial growth of the regular magnetic field due to the galactic
dynamo is saturated at some level, which is known as the equilibrium
magnetic-field strength ($B_{\rm eq}$) at which dynamo action
reaches equilibrium with turbulent dissipation. The equilibrium
magnetic-field strength is related to the equipartition field
strength ($B_*$):

\begin{equation}
{{\rho v^2} \over 2} = {{B^2_*} \over {8 \pi}} \label{equip}
\end{equation}
or

\begin{equation}
B_* = v \, \sqrt{4 \pi \rho}. \label{Beq}
\end{equation}

The more elaborated estimate by Shukurov (2004) takes into account
the intensity of dynamo action measured by $D_{\mathrm d}$:

\begin{equation}
B_{\rm eq} = B_* \, \sqrt{{D_{\mathrm d} \over {D_{\mathrm
{cr,\,d}}}} -1} \, .
 \label{eqmod}
\end{equation}
Using Eq.~(\ref{dynnum}) and $D_{\rm cr}=7$, we obtain

\begin{equation}
B_{\rm eq} = v \, \sqrt{4\pi \rho} \, \sqrt{1.28 {\left( h\Omega
\over v \right)^2} - 1} \, . \label{eqmf}
\end{equation}

The concept that $B_{\rm eq}$ is determined by $B_*$ is an
oversimplification. A deeper understanding of dynamo saturation
based on magnetic helicity conservation (e.g. Kleeorin et al. 2002)
yields a similar estimate for $B_{\rm eq}$.

We conclude that the timescale of dynamo growth and equilibrium
field strength are determined by different governing parameters. In
particular, if $v$ becomes larger (and the other parameters are
fixed), the dynamo timescale remains constant (Eq.~\ref{eq:ts_dd}),
while the equilibrium strength increases. On the other hand, the
dynamo number strongly decreases with increasing $v$
(Eq.~\ref{dynnum}) and may drop below the critical value.

\subsection{Timescale of regular magnetic fields in quasi-spherical
galaxies} \label{sec:dynamo_thick}

We estimate the dynamo timescales in spherical or ``puffy''
galaxies with $h/R > 0.1$. For a spheroid, $h \approx R$:

\begin{equation}
|D_{\mathrm s}| = R_\alpha R_\omega \simeq 9 \left({{R \Omega} \over
v}\right)^2,
\label{Dsp}
\end{equation}
while the dynamo growth rate is

\begin{equation}
\Gamma = |D_{\mathrm s}|^{2/3} {\beta \over {R^2}}
 \label{shgr}
 \end{equation}
(Sokoloff 2002). This yields the following estimate of the dynamo
timescale $t^{**}$ of a spheroid:

\begin{equation}
t^{**} = {3 \over {9^{2/3}}} \left({v \over {R \Omega}}\right)^{1/3}
{R \over {\Omega l}}.
 \label{eq:ts_sd}
\end{equation}
We note that the final term is the dynamo timescale for a disk $t^*$,
where $h$ is replaced by $R$. Assuming that $v = 10$ km s$^{-1}$ and $R
\Omega = 250$ km s$^{-1}$, we estimate that

\begin{equation}
t^{**} = 0.23 t^*(R),
 \label{crest}
\end{equation}
where $t^*(R)$ means that in Eq.~(\ref{eq:ts_dd}) $h$ is replaced by
$R$. Since $R$ is generally a few times larger than $h$, the timescales
for disk and spherical galaxies are more or less comparable.

The dynamo number of a quasi-spherical galaxy is calculated from the
ratio of Eq.~($\ref{eq:ts_dd}$) to Eq.~($\ref{Dsp}$), $|D_{\mathrm s}|
= |D_{\mathrm d}| (R/h)^2=10^3-10^4$, for $R/h=10$ and $|D_{\mathrm
d}|=10-100$ (e.g. Belved\'ere et al., 1998). The critical dynamo
number $|D_{\mathrm {cr,\,s}}|$ for a quasi-spherical body is
within the wide range between 300 and $5 \times 10^3$ (cf. Sokoloff
\& Shukurov (1990) and Sokoloff et al. (2008)), depending on the details of
the rotation curve and the spatial distribution of the
dynamo-governing parameters. For the above estimates, $|D_{\mathrm
s}| \approx 5000$. In this paper, we assume that $|D_{\mathrm
s}|>|D_{\mathrm {cr,\,s}}|$ and do not specify the suppression of
the mean-field dynamo in spherical bodies.

\section{Magnetic-field generation by the turbulent dynamo}
\label{sec:ts_td}

Magnetic-field generation by the turbulent (small-scale) dynamo in
galaxies requires neither large-scale rotation nor a disk, only
turbulence. It produces magnetic fields on scales comparable to the
basic scale of galactic turbulence on timescales far shorter than
that for the conventional mean-field galactic dynamo. The timescale
(e-folding time) of the small-scale dynamo is given by
\begin{equation}
  \tau = l/v,
  \label{eq:ts_td}
\end{equation}
(see Batchelor (1950), Landau \& Lifshitz (1959), and Kazantsev (1967)), while
the equilibrium strength $b$ remains comparable to $B_*$.

Considering that interstellar turbulence produces random
small-scale magnetic fields as well as regular large-scale ones, we
compare their strengths. We note that the term $\sqrt{D/D_{\rm cr}
-1}$ in Eq.~(\ref{eqmod}) is usually less than unity, so that the
ratio of small-scale and large-scale field strengths $b/B_{\rm eq}$
is expected to be slightly larger than unity. A ratio of $b/B_{\rm
eq} \simeq 1.7$ is estimated from observations (Ruzmaikin et al.
(1988), see however Beck et al. (2003) for a systematic bias imposed on
$b/B_{\rm eq}$). For the MW, this ratio, derived from the above
theoretical arguments, is

\begin{equation}
{b \over {B_{\rm eq}}} = {1 \over {\sqrt{1.28({{h \Omega} \over
v})^2 -1}}}\approx 1.9, \label{ratio}
\end{equation}
in good agreement with the observational value.

\section{Origin of seed magnetic fields} \label{sec:osmf}

The small-scale and large-scale dynamo mechanisms cannot explain the
origin of magnetic fields. Seed magnetic fields are required for the
dynamo mechanism to amplify and maintain the magnetic field themselves.
One possible origin of a seed field is the Biermann battery mechanism, which was
initially suggested in a cosmological context by Harrison (1970) and
applied to protogalaxies by Mishustin \& Ruzmaikin (1971) (see
Ruzmaikin et al. (1988) for a review). The typical magnetic-field
strength produced by the battery mechanism is estimated to be between
$B_{\rm seed} = 10^{-20}$ and $10^{-22}$~G.

In the context of a hierarchical structure formation scenario, the
Biermann battery is able to generate a field of order $10^{-18}$\,G
in a protogalactic cloud at $z\simeq 40$ (Pudritz \& Silk 1989;
Davies \& Widrow 2000). This theoretical prediction is consistent
with magnetohydrodynamical simulations of Population III star
formation by Xu et al. (2008). Magnetic fields of order $\simeq
10^{-18}$ G are generated predominantly by the Biermann effect, early
in the evolution of the halo ($z=40$) at low densities and large
spatial scales (400 pc).

Other possible mechanisms for generating magnetic seed fields are
kinetic plasma instabilities, the so-called Weibel instability
(Weibel 1959) in cosmological shocks (Schlickeiser \& Shukla 2003;
Medvedev et al. 2004) or cosmological perturbations (proposed by
Takahashi et al. (2006)) that may produce field strengths of $\sim
10^{-19}$~G (see also the review by Semikoz \& Sokoloff (2005)).

In summary, magnetic fields can be generated by various mechanisms
and during various stages of cosmological evolution. It is, however,
difficult to generate a regular magnetic field on galactic scales
that is of reasonable strength in the pre-recombination era when the
Universe was homogeneous and isotropic; the creation of
magnetic fields instead requires timescales and peculiar motions with
respect to the homogeneous and isotropic cosmological background
that are inconsistent with the assumptions of isotropy and homogeneity.

The variety of possible mechanisms for generating seed
magnetic fields increases with the development of spatial structures
in the Universe. The first magnetic fields that represented seeds of
galactic magnetic fields were probably created sometime between the
epoch of reionization and formation of galaxies ($10\la z\la 40$).

A mechanism creating small-scale magnetic fields with a vanishing
mean value is able to randomly generate a weak large-scale field
component of

\begin{equation}
B_{\rm s} = b \, N^{-1/2}
\label{seedloc}
\end{equation}
where $N$ is the number of turbulent cells in a galaxy. This is
sufficient to represent a seed for the large-scale galactic dynamo, as pointed
out by Ruzmaikin et al. (1988), Beck et al. (1996), and Subramanian
(1998). The key point is that the number of turbulent
cells is large but not enormous, much less then the Avogadro number $A=6
\times 10^{23}$, the typical number of particles in laboratory
statistical physics, where terms of order $N^{-1/2}$ can be
neglected.

Using $N= \pi R^2 2 h/(4 \pi l^3/3) \approx 8 \times 10^4$ for
a disk galaxy with $R=10$~kpc, $h=500$~pc and $l=100$~pc and
assuming that the small-scale magnetic field is on the equipartition
level, $b\approx B_*$, we obtain

\begin{equation}
B_{{\rm s}} \approx B_*/300. \label{seed glob}
\end{equation}
Hence, the large-scale dynamo has to amplify the seed field by only
a factor of about 300, much less than the upper limit of the
amplification factor of $10^{17}$ obtained from Eq.~(7), and the corresponding amplification timescale is then far shorter than the galaxy age.

The evolution of galactic magnetic fields initiated with a seed field
generated by the Biermann battery (in the epoch of protogalaxy formation of
protogalaxies), which was followed by field amplification by the turbulent
small-scale dynamo, and further amplification and ordering by the
large-scale dynamo. A deeper understanding of the process would also
require inclusion of the magneto-rotational instability (Kitchatinov
\& R\"udiger 2004), although this is beyond the scope of the present
paper.

\section{Formation and evolution of disk galaxies}
\label{sec:edg}

We review results of recent simulations of the
hierarchical merging of dark halos, and the formation of protogalaxies and
their evolution (no concise review of the recent numerical results
exists so far). We attempt to identify the mechanisms
responsible for generating and amplifying turbulent and
regular magnetic fields in the halo and disk of galaxies, in developing
a simple model for the evolution of magnetic fields and determining
the earliest epochs at which large-scale regular magnetic fields are
formed and amplified in disk galaxies, of the type considered in the next Section.
The formation and evolution of regular large-scale magnetic fields
is related intimately to the formation and evolution of disks in
galaxies in terms of geometrical and physical parameters such as the radius
$R$, half-thickness $h$, and angular momentum $\Omega$ of the galaxy,
turbulence velocity $v$, turbulence scale $l$, and density $\rho$ of
the gas. These parameters evolve differently for different galaxy
types and sizes.

We adopt the standard hierarchical cold dark matter cosmology
in which the cosmic structures and galaxies assemble by the merging of
small dark matter halos (Kauffmann et al. 1993; Baugh et al. 1996).
The first very massive stars ($\ga 100$\,M$_{\sun}$) are predicted
to have formed in dark matter halos of virial mass $\sim
10^6$\,M$_{\sun}$ at redshifts $z\la 20$ (e.g. Tegmark et al. 1997;
Omukai \& Palla 2003). The dark matter halos merged, assembling the
first protogalaxies of masses $\ga 5\times10^7 M_{\sun}$ at $z\ga
10$ (Press \& Schechter 1974; Rees \& Ostriker 1977). The gas and
stellar components of protogalaxies may have had a quasi-spherical
or disk-like distribution, depending on the initial conditions of
simulations (Brook et al. 2004; Wise \& Abel 2007; Greif et al.
2008). The first galaxies were able to retain photo-heated gas and
maintain self-regulated star formation (e.g. Ricotti et al. 2008),
and developed into disk galaxies.

In this scenario, we identify two main cosmological epochs at which
magnetic fields can be generated and amplified by different dynamo
mechanisms. In the first epoch of the \emph{virilization and merging
of dark matter halos}, which occurred between $35\ga z \ga 10$, the
mass of protogalaxies assembled via gas accretion onto the halos and
minor/major mergers of dark matter halos. High-resolution
simulations of early pre-galactic halos (Wise \& Abel 2007; Greif et
al. 2008), which include primordial gas cooling and mass accretion
history, showed that significant turbulence was generated in all
cosmological halos during thermal virilization. The gas accreted
from the IGM was shock-heated to the virial temperature. This mode
of accretion (hot accretion) worked effectively at high redshifts in
low-mass halos and generated turbulence on scales comparable with the
size of the infalling gas. As radiative cooling became efficient,
the gas attempted to virialize by increasing its kinetic (turbulent)
energy, which was achieved by radial infall and turbulent motions.
The turbulence became supersonic with Mach numbers ranging from 1 to
3 (Wise \& Abel 2007).

In this mode (cold accretion), massive filaments could form molecules,
which allowed efficient cooling of the filaments and their flow at
high velocities ($\sim20$ km s$^{-1}$) towards the center of the
protogalaxy (Greif et al. 2008). Inflows of cold gas were supersonic
with Mach numbers $\sim10$ and reached the central part of the
galaxy, generating significant turbulence with Mach numbers ranging
from 1 to 5.

Virial turbulence may have been most important in halos with
efficient radiative cooling rates, for example in halos of masses
below $10^{12}$ M$_{\sun}$ that could be cooled effectively by
Ly$\alpha$ emission (Birnboim \& Dekel 2003; Dekel \& Birnboim
2006). For low-mass protogalaxies, the largest driving scale of
turbulence was $\sim 200$ pc (1/3 of the virial radius) and the
r.m.s. velocities was around $20$ km s$^{-1}$ (John Wise, private
communication). The dominant role of virial turbulence played an
important role in star formation in the regions with density
enhancements and the amplification of magnetic fields by means of
the turbulent dynamo (see Sect.~\ref{subsec:threephasemodel}).

Mergers could also have generated turbulence that started on length scales of
merging halos and developed to small scales by means of turbulence cascades.
Another driver of turbulence was the Kelvin-Helmholtz instability
between bulk flows and virialized multi-phase gas (Takizawa 2005),
which may have occurred during minor mergers. Simulations of
protogalaxies in the redshift range between 35 and 15 demonstrated that
the turbulent Mach number depends on merger history. The Mach number
was supersonic, reaching values between 2 to 4 in the case of two
subsequent major mergers. Before the central collapse occurred in
these halos, turbulence was mostly subsonic ($<1$; Wise et al.
2007). Major mergers of halos resulted in high turbulence Mach
numbers and hence high turbulent velocities.

The beginning of the second epoch manifested itself by \emph{the
formation of an extended large-scale disk} in the first galaxies.
Highly resolved simulations showed that the first galaxies were born
at redshifts $z\simeq 10$, after the atomic cooling criterion was
fulfilled and they were able to retain the photo-heated gas (Greif
et al. 2008). Brook et al. (2004) demonstrated that the thick
stellar disk was created during an epoch of multiple mergers of
gas-rich halos, which were abundant at high redshifts. The population
of stars forming during the merging period formed the thick disk.
The angular momentum of a significant fraction of the accreted gas
(i.e. the angular momentum of the protogalactic halo) resulted in
the rotation and flattening of the formed galaxy. The forming thick
gas disk was dynamically hot, resulting in the formation of
high-velocity stars in the thick disk. After the epoch of violent
merger events, the gas was accreted and formed a smooth thin disk.

Recent models of disk formation have included hierarchical dark matter
cosmology, gas cooling, star formation, and supernova (SN) feedback
(Governato et al. 2007; Stringer \& Benson 2007) and benefitted from
the increased resolution of numerical simulations (Mayer et al.
2008). High mass resolution and spatial resolution are important in
simulating large MW-type disks (Mayer et al. 2008). The formation
of disks at high redshifts ($z>2$) can be simulated with relatively
high spatial and mass resolutions (Governato et al. 2007). The disk
started forming in galaxies with $M \ga 10^{11}$ M$_{\sun}$
immediately after the last gas-rich major merger with substantial
stellar feedback (Elmegreen et al. 2005; Springel \& Hernquist 2005)
at $z_{\mathrm{llm}}\sim 2$. Preserving the angular momentum from
the merger, the gas cooled and formed the large-scale exponential
disk. This effect was more apparent in numerical runs with stronger
feedback and higher resolution. After the stellar disk had formed, it
remained fairly stable against merger events provided there was
sufficient gas to form stars (Springel \& Hernquist 2005). As discussed by Kaufmann et al. (2007), internal and
external UV background radiation would prevent cooling below $\sim
10^4$ K in the warm galactic gas by providing pressure support
dynamically comparable to angular momentum support. They showed that pressure support was less important in
large halos ($\ga10^9$\,M$_{\sun}$). This created a
tendency to form a thin disk in massive, isolated galaxies, while
their less massive counterparts (dwarf galaxies) remained
spheroidal, puffy systems with more gas and less efficient star
formation than in the larger MW-type galaxies. The disk could settle
down (and survive until the present time) in massive halos after the
epoch of gas-rich mergers, even at redshifts $z\ga5$ (Lucio Mayer,
private communication). Future higher-resolution numerical
simulations will allow to trace the disk formation back to
$z\sim10$.

The advanced disk-formation model presented by Stringer \& Benson
(2007) showed that the resulting disks of the isolated galaxy (no
merger) and the galaxy formed through minor mergers had very
similar evolutionary behaviors. The disk component was not destroyed
and evolved into a disk galaxy. The morphology of the formed disk
galaxies at $z\la1.5$ remained unchanged in the case of no merger or
only minor mergers, while in the case of frequent minor mergers or a
major merger the disk galaxy could evolve into a disk galaxy having
a spheroidal or bulge component, or into an elliptical galaxy
without disk (see Sect.~\ref{subsec:isfm}).

If a late major merger destroyed the stellar disk at later epochs
($z\sim1$), the gaseous and stellar disks could recover,
if there was sufficient gas remaining to form new stars.

It is generally believed that, in the hierarchical cold dark matter
Universe, the massive disk galaxies were formed first, while the
majority of present-day galaxies along the Hubble sequence such as
elliptical galaxies and disk galaxies with bulges formed at later
epochs by the merging of giant disk galaxies (Kauffmann et al.
1993; Baugh et al. 1996). The existence of massive galaxies at high
redshifts is verified by the detections of disk-like or clumpy
disk galaxies between $1.4<z<3$ from deep near-infrared imaging
(Labb\'e et al. 2003) and at $z\sim5.5$ (Elmegreen et al. 2007).
Observations of luminous star-forming galaxies at high angular
resolution showed that large and massive rotating protodisk galaxies
were present already at $z \sim 2$ to $3$ (Genzel et al. 2006).

\section{Evolution of magnetic fields in galaxies}
\label{sec:gmfdg}

\subsection{Three-phase model}
\label{subsec:threephasemodel}

In the hierarchical formation scenario, we identify three main
phases of magnetic-field evolution in galaxies. In the first phase,
the seed magnetic fields of order $\simeq 10^{-18}$\,G (Pudritz \&
Silk 1989; Davies \& Widrow 2000; Xu et al. 2008) were generated in
dark matter halos by \emph{the Biermann battery mechanism}, well
before the formation of first massive stars at $z\sim20$.

The second phase started at $z\sim20$ (or $t\sim0.5$\,Gyr) when the
merging of halos and virialization generated turbulence in the halo.
According to Wise \& Abel (2007), this epoch was dominated by
turbulence generated during the thermal virialization of halos.
There the \emph{turbulent dynamo} could effectively amplify the seed
magnetic field of halos to the equilibrium level
(Eq.~\ref{ratio}), $B_{\mathrm s}\sim 20~\mu$G, on timescales given
by Eq.~(\ref{eq:ts_td}).

In the third phase, the \emph{mean-field dynamo} mechanism started
acting in the newly formed galaxies at $z\sim10$. The first gas-rich
massive galaxies ($\ga10^9$ M$_{\sun}$) formed extended thin disks
($h/R\la0.1$) at $z_{\rm disk}=10$ after major merger events, as
discussed in Sect.~\ref{sec:edg}. A weak, large-scale magnetic-field
component was generated in the disk from small-scale magnetic
fields of the halo (Eq.~\ref{seedloc}), amplified to the equilibrium
level in the second phase. Then, the ``disk'' mean-field dynamo
amplified and ordered weak, regular magnetic fields on timescales given by Eq.~(\ref{eq:ts_dd}) and Eq.~(\ref{eq:ts_o}),
respectively. If the formed disk was thick ($h/R\ga0.1$)
or the disk had not formed, the ``quasi-spherical'' mean-field
dynamo acting in puffy objects would be switched on. It would amplify the field
to the equilibrium level and order the magnetic fields on scales
of a few kiloparsecs and on timescales given by
Eq.~(\ref{eq:ts_sd}) and Eq.~(\ref{eq:ts_o}) ($h\approx R$),
respectively. If a thin disk in puffy objects had been formed at
later epochs, the ``disk'' mean-field dynamo would have dominated and amplified
the regular fields at the equilibrium level. This scenario is more
appropriate for dwarf galaxies ($\la10^9$ M$_{\sun}$) that could not
maintain the disk at early epochs because of strong UV background
radiation, and may have formed a disk at later epochs ($z\sim 2$)
when UV radiation was weak and allowed efficient radiative cooling.

If the disk was destroyed by a late gas-rich, major merger event, the
regular magnetic field must have dissipated and increased the amplitude of the
turbulent magnetic field. During the disk recovery, the regular
magnetic field must have been re-generated and amplified back to its
equilibrium level (Eq.~\ref{eqmf}). The seed field for the
large-scale dynamo is taken from Eq.~(\ref{seedloc}) using $N$
according to the appropriate galaxy radius.

The sketch of the evolution in the amplitude and ordering scale
(Eq.~\ref{eq:ts_o}) of magnetic fields in dwarf and spiral galaxies
is presented in Figs.~1-3.

\subsection{Assumptions of the model}
\label{subsec:am}

We assumed an average gas density of a halo (adiabatic model) within
its virial radius of about 600\,pc at $z=15$ of $\approx 6\times
10^{-24}$ g\,cm$^{-3}$(Wise \& Abel 2007). The gas densities of
nearby disk galaxies are calculated to be in the range of between $
10^{-24}$~g\,cm$^{-3}$ and $15\times 10^{-24}$~g\,cm$^{-3}$ for a random
magnetic-field strength in the range $\sim(2-10)~\mu$G (Beck et al. 1996).
For simplicity, we assume the same gas density for halo- and
disk-dominated galaxies and that their gas density and angular
rotation velocity are unchanged during the evolution, i.e. $\rho =
10^{-23}$ g\,cm$^{-3}$ and
$\Omega=20\,\mathrm{km}\,\mathrm{s}^{-1}\,\mathrm{kpc}^{-1}$.

\emph{Halos}. We assume that virial turbulence dominates during the
epoch of merging and virialization of dark matter halos (Wise \&
Abel 2007). The typical turbulence velocity and the largest scale of
turbulence, 20\,km\,s$^{-1}$ and 200\,pc in halos of masses $\la
10^{8}$\,M$_{\sun}$ (John Wise, private communications) and are
adopted for halos of masses $\la 10^{12}$\,M$_{\sun}$.

\emph{Galaxies}. For massive and MW-type disk galaxies ($\ga
10^{10}$\,M$_{\sun}$), we assume that (a) the gaseous and stellar
disks formed at $z_{\rm disk}\approx 10$; (b) the turbulence scale
and turbulence velocity were driven by SN explosions and had values
close to those observed in present-day disk galaxies such as the MW,
$l=100$\,pc and $v=10$\,km\,s$^{-1}$; (c) the scale height of the
disk, $h=500$\,pc, remained unchanged as demonstrated by N-body
simulations of late-type disk galaxies at $z<1$ (Brook et al. 2006),
while its scale length in the plane decreased according to
$R_z=R\,(1+z)^{-0.45}$ (Trujillo et al. 2006), where
$R$(z=0)\,$=10$\,kpc. This is in reasonably good agreement with the
observed disk thickness ratio of $h/R\sim0.08$ at $z\sim1$
(Reshetnikov et al. 2003) and $h/R\sim0.15$ at higher redshifts
(Elmegreen et al. 2005).

For typical low-mass puffy galaxies ($\la10^{10}$ M$_{\sun}$), we
adopt $R=3$\,kpc, $h \approx R=3$\,kpc, and $\Omega=20$\,km s$^{-1}$
kpc$^{-1}$. The ordering timescale of the field in these galaxies is
calculated from Eq.~(\ref{geommean}) using $h \approx R$. We assume
that the ``quasi-spherical'', mean-field dynamo amplified the regular
magnetic field in puffy galaxies ($h/R > 0.1$), while the ``disk''
mean-field dynamo was effective in galaxies with a thin disk ($h/R < 0.1$).

\subsection{Isolated disk galaxies}
\label{subsec:idg}

\begin{figure}
\center
\includegraphics[angle=-90,width=8cm]{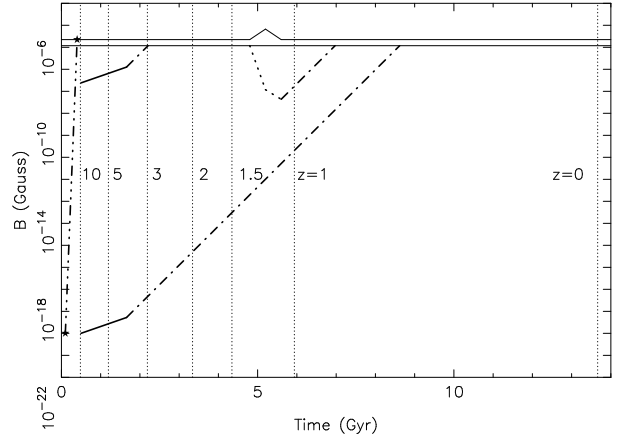}
    \caption{Evolution of magnetic fields in MW-type disk galaxies:
    magnetic-field strength versus cosmic epoch. Evolution of the small-scale magnetic
    field generated by the turbulent dynamo (thick dashed-dot-dot-dot line) and the
    large-scale magnetic
    field generated by the large-scale mean-field dynamo in quasi-spherical galaxies (thick
    solid line)
    or in thin-disk galaxies (thick dashed-dot-dashed). Dissipation of the field because
    of a major merger event is presented by a dotted line. The lower curve traces the
    evolution of
    regular magnetic fields generated by the pure large-scale dynamo mechanism (no
    amplification by
    the turbulent dynamo). The two horizontal thin solid lines represent the equipartition
    and
    equilibrium magnetic-field strengths (upper and lower lines, respectively). The vertical
    thin
    dotted lines indicate redshifts from 0 to 10.}
 \label{fig:mfe1}
\end{figure}

\begin{figure}
\center
\includegraphics[angle=-90,width=8cm]{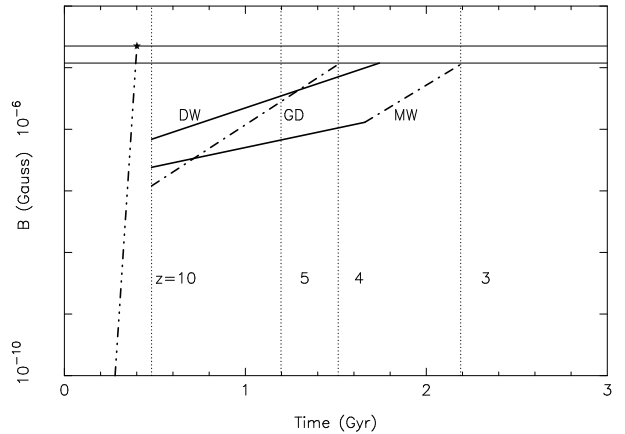}
    \caption{Evolution in the magnetic-field strength of dwarf galaxies (DW), MW-type, and giant galaxies (GD). The meanings of different line types are the same as in Fig. ~\ref{fig:mfe1}. }
 \label{fig:mfe2}
\end{figure}

We consider the evolution of magnetic fields in giant disk galaxies
(called GD hereafter; $R$(z=0)\,$=20$\,kpc), MW-type galaxies (10\,kpc), and
dwarf galaxies (called DW hereafter; $R$(z=0)\,$=3$\,kpc). We assume
that the seeds of turbulent magnetic fields of strength $\approx 10^{-18}$\,G existed in the protogalaxies of present disk
galaxies, at $z \approx 35$. Virial turbulence could amplify
turbulent magnetic fields in merging dark-matter haloes via the
small-scale dynamo from $10^{-18}$\,G seed field strength to reach
the equipartition field strength of $2.2 \times 10^{-5}$\,G at
$z\simeq 11$ (0.4\,Gyr) within a short period of time, $\approx 3
\times 10^8$\,yr (Eq.~\ref{eq:ts_td}; see Figs.~\ref{fig:mfe1} and
\ref{fig:mfe2}). For simplicity, we consider that all types of
galaxies reached the equipartition level of turbulent fields at the
same epoch, $z \simeq 11$.

\begin{figure}
\center
\includegraphics[angle=-90,width=8cm]{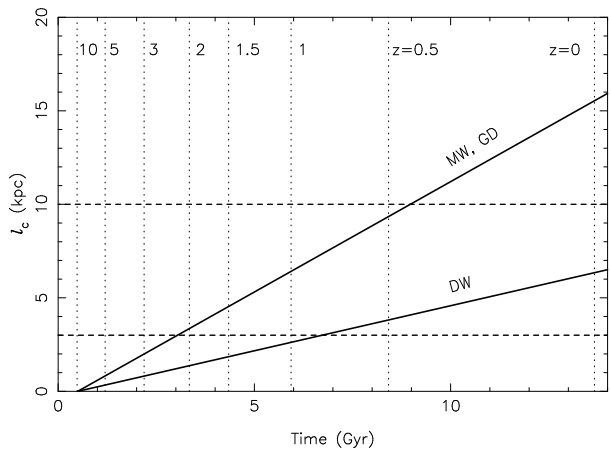}
    \caption{Scale of ordering of regular magnetic fields with cosmological epoch.
    The evolution of the ordering magnetic fields is shown for
    dwarf galaxies (DW; bottom line) and disk galaxies (MW and GD; top line). }
 \label{fig:mfe3}
\end{figure}

\emph{MW-like galaxies.} A sketch of the evolution in the magnetic
fields of isolated MW-type galaxies with $M_{\rm MW}\sim
10^{11}M_{\sun}$ is shown in Figs.~\ref{fig:mfe1} and
\ref{fig:mfe2}. At $z_{\rm disk} \approx 10$, the disk was formed and
evolved in isolation (no major merger) to the present time. The size
of the galaxy was $R_{10}(z=10)=R\,(1+z)^{-0.45}=3.4$\,kpc and
$h/R_{10}=0.14>0.1$, and the ``quasi-spherical'' mean-field dynamo
amplified the field until $z\approx 4$, at which time, $h/R$ became less
then 0.1. The ``disk'' mean-field dynamo then became significantly more important and
amplified the regular large-scale field within $\approx 1.5$\,Gyr,
and reached its equilibrium state at $z\approx3$
(Fig.~\ref{fig:mfe1}). At this epoch, the regular field was ordered
on a scale of a few kiloparsecs (Fig.~\ref{fig:mfe3}) and finally
reached a coherence scale similar to the size of the MW (10\,kpc) at
$z\approx 0.4$.

The earliest regular magnetic fields of equipartition strength were
generated in $t_{\rm d,global}\approx 1.4$~Gyr after disk formation
(Fig.~\ref{fig:mfe1}), while the ordering of magnetic fields on a length scale similar to that of a MW-type galaxy was complete after $\approx 9$~Gyr
(Fig.~\ref{fig:mfe3}). Hence, present-day, MW-type galaxies are
expected to host fully ordered regular fields.

This evolutionary scenario also corresponds to the evolution of a
disk galaxy that experiences multiple minor mergers that do not destroy its
disk. Multiple and even multiple frequent minor mergers may have caused local distortion of the disk, resulting in a high
star-formation rate that shifted the start of large-scale dynamo
action to a later epoch. This will be discussed in more detail in
the next Section.

If amplification by the turbulent dynamo is not considered, the
mean-field, large-scale dynamo would have to amplify a seed field of
$10^{-18}$~G, and the time to reach the equilibrium strength of the
regular magnetic field would be far longer ($\approx 9$~Gyr, until
a redshift of $z \approx 0.3$) (Fig.~\ref{fig:mfe1}). However, this
scenario is unrealistic because sufficient
turbulence exists in young galaxies to drive the turbulent dynamo. Even
without turbulence, another mechanism may enhance the seed field
strength by a factor of $\ga10^4$: the dissipative collapse into a
disk (Lesch \& Chiba 1995) at $z\approx10$. This would enable
the equipartition level to be reached earlier, within $\approx 7.5$ Gyr.
In case of turbulence, this process is not required because the
turbulent field is already strong at this redshift. However, if the
collapse of the halo occurs at $z\ga10$, when the field amplitude
remains smaller than the equipartition level (see
Fig.~\ref{fig:mfe1}), this mechanism may further accelerate the
field amplification by the turbulent dynamo. In the following, we
only consider amplification of the seed field by the turbulent
dynamo.

\emph{Giant galaxies}. Late-type disk galaxies of a disk scale
length $\ga~10$~kpc are rare as verified by analyses of I-band images
(Courteau et al. 2007). In these giant disk galaxies ($M_{\rm
GD}\sim 10^{12}M_{\sun}$), the ratio $h/R_{10}(z=10)=0.5/6.8<0.1$
implies that the ``disk'' mean-field dynamo has already been switched on at
$z\approx 10$ (Fig.~\ref{fig:mfe2}), had amplified the regular magnetic
field within only $\approx 1$~Gyr, and reached the equilibrium
state already at $z\approx 4$. However, as shown in
Fig.~\ref{fig:mfe3}, the regular magnetic fields can be ordered only
on scales of 15~kpc until the present time. Hence, the mean-field
dynamo cannot generate coherent magnetic fields over the size of the
disks of giant galaxies ($R\ga15$\,kpc).

\emph{Dwarf galaxies}. We assume that, in low-mass halos, the
turbulent magnetic fields evolve in the same way as in their massive
counterparts. N-body/smoothed particle hydrodynamics simulations
demonstrated that low-mass galaxies did not at first form thin disks
because of the ionizing UV background radiation, which prevented
cooling of the warm galactic gas (Kaufmann et al. 2007). They were
more spheroidal, puffy systems ($h/R\approx 0.3$) of rotational
velocities $\approx 40$ km s$^{-1}$ and mass $M\approx 10^{10}$
M$_{\sun}$. In thick galaxies, the ``quasi-spherical'' dynamo could
generate and amplify the regular magnetic field to the equilibrium
field strength within $\approx 1.5$~Gyr (at $z \approx 3.5$;
Fig.~\ref{fig:mfe2}). At $z\approx2$, the UV background intensity
decreases (Bianchi et al. 2001), resulting in the formation of a
thin disk in a small galaxy that preserves the strength and
ordering of the existing regular field. Although the ordering
timescale of dwarf galaxies is longer (Fig.~\ref{fig:mfe3}), fully
coherent regular magnetic fields were generated at earlier epochs
($z\approx 1$) because of the smaller sizes of dwarf galaxies. \\

Due to the hierarchical clustering, it is expected that more massive
protogalaxies formed at later epochs. In this scenario, the first
disks should have formed in MW-type galaxies (see
Fig.~\ref{fig:mfe2}) and, hence, the time to reach magnetic-field
strengths of $\sim 10^{-6}$~G in MW-type and giant galaxies would be
comparable, while the ordering of the regular fields would have been
completed earlier in MW-type galaxies. If dwarf galaxies formed
earlier ($z\ga 10$), the regular field would have reached the
equilibrium magnetic-field strength earlier and would have been
ordered even earlier, at $z\approx 1$ (Fig.~\ref{fig:mfe3}).

\subsection{Influence of star formation and mergers}
\label{subsec:isfm}

\emph{Star formation}. The evolution of regular magnetic fields
depends on the star-formation rate (SFR) and the disk parameters. Models of disk formation demonstrated that star formation is a
fundamental parameter. It can be triggered in isolated galaxies by gravitational instability, in interacting galaxies by minor and
major mergers, by tidal forces, leading to the compression of the
gas, and by interactions of high velocity HI diffuse clouds (e.g.
Kennicutt et al. 1987; Combes 2005). Star formation is more
efficient at high redshifts because of the high merging rate and
more gas at $z\simeq 1.3$ (Lefevre et al. 2000; Ryan et al. 2008),
and in more massive disk galaxies (Governato et al. 2007).

Major mergers of gas-rich galaxies may trigger high SFR,
causing high velocity turbulence of the ionized gas, which in turn
can suppress the mean-field dynamo in the thick disk if the dynamo
number is $|D_{\mathrm d}|<D_{\mathrm c}\approx100-300$ and in the
thin disk if $|D_{\mathrm c}|<D_{\mathrm c}\approx7$ (see
Sect.~\ref{sec:dynamo_thick} and \ref{subsec:dynamo_thin}). The
latter restriction places an upper limit on the turbulent velocity of
the gas:

\begin{equation}
  \label{eq:v}
  v \la 1.1\, \Omega\, h = 11 \,\,\mathrm{km\, sec}^{-1},
\end{equation}
where $\Omega=20\,\mathrm{km}\,\mathrm{s}^{-1}\,\mathrm{kpc}^{-1}$
and $h=0.5$\,kpc. The turbulent velocity (or characteristic velocity
dispersion of the gas) and the SFR correlate positively for nearby
galaxies (Dib et al. 2006). Using this relation and Eq.~(\ref{eq:v}),
one derives an upper limit of $\mathrm{SFR}\la20$ (in units of
$M_{\sun}\,{\mathrm{yr}^{-1}}$) up to which the action of the
large-scale dynamo is possible. This indicates that the merging
history of the galaxy had an important influence on the
evolution of magnetic fields. In this scenario, regular fields were generated primarily at later epochs, $z<4$ (see
Fig.~\ref{fig:mfe1}).

\emph{Major mergers} were rare but could alter the morphology of a
spiral galaxy and destroy its regular magnetic field. After the
merger, there were two possibilities, firstly, to form a disk galaxy
with spheroidal or bulge component, or secondly, to form an
elliptical galaxy without a disk. The SFR during the major merger
($\la 2\times10^9$ yr) was $\sim50$ times higher than in the case of
an isolated galaxy (di Matteo et al. 2007; Bournaud et al. 2007).
This high SFR produced an increase in total magnetic-field
strength as shown by the peak of the equipartition magnetic-field
strength at about 5~Gyr in Fig.~\ref{fig:mfe1}.

Disk galaxies, which survived after a gas-rich major merger, formed a
thin-disk component immediately after the merger event and needed
$\approx 1.5$ Gyr to amplify the regular magnetic field to the
equilibrium level by the mean-field dynamo (Fig.~\ref{fig:mfe1}) and
$\approx 8$ Gyr to generate a coherent magnetic field on the length scale of the galaxy size (Fig.~\ref{fig:mfe3}).

\emph{Minor mergers} were more frequent and may also have altered
the morphology (spiral into elliptical, spiral to spheroid),
increased the size and thickness of the disk, and controlled the star
formation rate (density, turbulence velocity). Simulations of the
formation of thick disks by minor mergers of satellites with a thin
disk showed that mergers with 10--20\% mass of the mass of the host
galaxy resulted in the formation of thick disks. The disk was not
destroyed but considerably heated up, tilted, and flared (Villalobos
et al. 2008). The scale length increased slightly with respect to
the original host disk, while the scale height became up to four
times larger, depending on the inclination of the satellite's orbit. A larger disk height and radius increased both the dynamo timescale
(Eq.~\ref{eq:ts_dd}) and the ordering timescale
(Eq.~\ref{geommean}), and shifted the formation of regular
magnetic fields to later epochs.

\emph{Multiple minor mergers}. Depending on the mass ratio of
galaxies and the number of minor mergers, the disk could have been
preserved, forming a spheroidal component, or destroyed,
forming multiple spheroids or an elliptical galaxy (Bournaud et al.
2007). The disk-formation model presented by Stringer \& Benson
(2007) illustrated that the resulting disk of an isolated galaxy and a galaxy formed by minor mergers have a very similar
evolutionary behaviors in which the main governing parameter is the infalling gas supplied from the hot halo.

\section{Discussion}
\label{sec:discussion}

\subsection{Observational tools}

The radio -- far-infrared correlation provides a powerful tool for detecting star formation in distant galaxies with the help of ground-based
radio observations. However, its validity depends critically on the
magnetic-field strength coupled with the interstellar gas
density and, hence the star-formation rate (Helou \& Bicay 1993;
Lisenfeld et al. 1996; Niklas \& Beck 1997). In galaxies with recent
starbursts, the radio continuum emission can be deficient with
respect to the far-infrared emission (Roussel et al. 2003). Either
the cosmic rays or the magnetic field or both are generated with
some time delay with respect to the dust-heating radiation field.

The total magnetic field can be measured by the observed total power
radio emission, corrected for the thermal fraction of a galaxy. The
regular magnetic field can be traced by polarized synchrotron
emission and by Faraday rotation.  While Faraday rotation is an
unambiguous signature of coherent regular fields, polarized emission
can also emerge from ``anisotropic'' fields, which are the result of
compressing or shearing of isotropic turbulent fields. This process
has been observed in several galaxies of the Virgo cluster (Vollmer
et al. 2007; Wezgowiec et al. 2007).

\subsection{Comparison with observations of galaxies}

Regular magnetic fields, as traced by polarized synchrotron emission
and Faraday rotation, were detected in the disks of nearby spiral
galaxies, with spiral patterns and amplitudes of up to 15 $\mu$G,
and were generally strongest in interarm regions (Krause 1990; Beck 2005).
Mean-field dynamo theory, in spite of its simplifications,
reproduces successfully the basic spiral pattern of magnetic fields
observed in nearby spiral galaxies as well as in barred galaxies
(Beck et al. 1996; Kulsrud 1999; R\"udiger \& Hollerbach 2004; Moss
et al. 2007). The prominent example is M\,31, which hosts a coherent
axisymmetric field within the emission ``ring'' at about 10~kpc
radius (Fletcher et al. 2004). However, regular fields with a
coherence scale similar to the galaxy size are rare among nearby
galaxies.

In contrast to the turbulent dynamo, the mean-field dynamo needs far
more time to generate a regular field. Fields with a
coherence length of about 1~kpc can be expected after 1--2~Gyr, at
$z\simeq 3-4$ (Fig.~\ref{fig:mfe3}), but the generation time for
regular fields with a coherence length of the galaxy size is several
times longer and comparable to the galaxy age (Fig.~\ref{fig:mfe3}).
The ordering timescale for a 10~kpc coherence scale such as that in
M\,31 is about 10~Gyr (Fig.~\ref{fig:mfe3}). The coherent field of
M\,31 is an indication that this galaxy has not experienced from a major merger during the past 10~Gyr.

Most other nearby galaxies observed so far have more complicated
field patterns than in M\,31 (Beck 2005). The long timescale required for
field ordering (Fig.~\ref{fig:mfe3}) is consistent with these
observations: most galaxies probably did not yet reach full field
coherence, either due to their size or due to major merging events.
The lifetime of galaxies larger than 15~kpc is insufficient to
generate a fully coherent field. If a major merger in a MW-type
galaxy occurred less than about 5~Gyr ago, the recovery of the fully
coherent regular field is not yet completed.

The results of this paper suggest a correlation between coherence
scale and galaxy size. The only direct indication of a regular
field in a distant galaxy was derived from Faraday rotation in an
intervening galaxy at z=0.395 measured against a background quasar
(Kronberg et al. 1992). Since the pattern exhibits reversals and hence is incoherent on the galaxy scale, this result is consistent with our
results. Indirect evidence of regular fields in foreground
spiral systems beyond $z\simeq 1$ has been provided by Faraday rotation
probes of distant radio sources (Kronberg et al. 2008; Bernet et al.
2008). They found that the scatter in the intrinsic RMs of spirals is
$\sim 20-50$ rad m$^{-2}$, which implies a coherence of the field on
scales of about half the galaxy size (in the case of no field
reversals). This coherence scale is consistent with the ordering
scale of magnetic fields derived in this work (see
Fig.~\ref{fig:mfe3}) for MW-type galaxies in the redshift range $z\approx1-2$.

In the LMC, another axisymmetric field was detected, which was
found to have properties conflicting with dynamo theory (Gaensler et al. 2005). Since the LMC has a disk, although this rotates slowly, the mean-field dynamo
can indeed operate and create a global coherent field on a scale of
4~kpc (radius of the LMC) within about 10~Gyr (Fig.~\ref{fig:mfe3}).
Hence, dwarf galaxies are prime candidates for hosting fully coherent
regular fields. On the other hand, dwarf galaxies are affected more significantly than larger galaxies by interactions. Simulations indicate that the LMC may have passed
through the plane of the Milky Way more than 2~Gyr ago, or that it is on its way to a first passage through the MW (Kallivayalil et al.
(2006), Piatek et al. (2008), and references therein). Since the regular
field would be severely distorted during each passage, the
observation of a regular field supports the idea of no passages in the past.

M\,82 is the prototypical and most nearby starburst galaxy, excited
by a tidal interaction with M\,81. Its total magnetic field is
strong, about $50~\mu$G (Klein et al. 1988), and probably in
equipartition with the other components of the wind outflow. The
radio emission from the galaxy disk appears to be almost completely
depolarized (Reuter et al. 1994). The radio halo is highly polarized
but Faraday rotation is small, indicating anisotropic turbulent
fields that are sheared and compressed in the rapidly moving gas outflow.
This confirms our model of a strong and dominant turbulent field
in starburst galaxies (Fig.~\ref{fig:mfe1}). Observations of higher frequency, however, also detected polarized
emission in the disk (Wielebinski 2006), indicating that some
regular disk field may have survived the tidal interaction, which
might not have been the case after a galaxy merger. To search for regular fields in
the disks and halos of starburst galaxies, new observations of higher resolution are required.

From Zeeman splitting in the HI absorption line against, a background
quasar (Wolfe et al. 2008) a $84 \pm 9~\mu$G field with a coherence length of
$\ge200$~pc was discovered in a low-metallicity galaxy at $z=0.692$. A turbulent dynamo can generate such a strong
turbulent field in a starburst galaxy triggered by a major merger
(see Fig.~\ref{fig:mfe1}). However, this field cannot produce a
Zeeman signal because of line-of-sight integration through
the whole galaxy, unless the observed field is
located in a single dense gas cloud or filament. Alternatively, the
observed field could originate in a shear or compression layer between interacting
galaxies, although verification of this possiblity is beyond the scope of this paper.

\subsection{Detection of synchrotron emission from distant galaxies with SKA}
\label{sec:dpdgws}

One major objective of SKA observations will be the detection of radio continuum
emission from distant galaxies to study the star-formation history
in the early Universe (van der Hulst et al. 2004). The tight radio -- far-infrared correlation in galaxies implies that radio synchrotron
emission is an excellent tracer of star formation in galaxies
(Condon et al. 1991), at least to distances of $z\simeq3$ (Seymour
et al. 2008). However, its application to even higher redshifts depends crucially on the existence of magnetic fields at the equipartition level with turbulent gas motions.

The timescale for field generation in young galaxies constrains the
largest distance to which SKA can detect star-forming
galaxies in synchrotron emission. We have shown in this paper that,
even in the case of very weak seed fields of $10^{-18}$~G, the turbulent
small-scale dynamo, with the help of virial turbulence, can amplify
turbulent fields to the level of equipartition with turbulent
energy density within $\simeq 3\times10^8$ years; strong
fields should therefore exist in all star-forming galaxies at
$z\simeq10$ (Fig.~\ref{fig:mfe1}) and the radio -- far-infrared
correlation should be valid for $z\la10$. Hence, radio continuum
emission can be used as a tracer of star formation in the early
Universe.

Star-formation rate and total magnetic-field strength are related
nonlinearly (Niklas \& Beck 1997; Chy\.zy et al. 2007; Krause 2008),
so that young starburst galaxies should have much stronger fields
and a higher radio -- far-infrared ratio than normal galaxies. The
SKA and its pathfinder telescopes will investigate this relation in
more detail.

The SKA will detect radio continuum emission from spiral galaxies
similar to the MW to $z\simeq3$ and to even larger distances for
more massive and starburst galaxies. The prediction from dynamo
models presented in this paper can be tested with the help of
polarization and Faraday rotation data from the SKA radio continuum
and RM surveys (Gaensler et al. 2004). Systematic observations of
galaxies at $z=1-5$ are expected to envisage the early stages of
galactic magnetic-field evolution.

The first regular magnetic fields of contemporary disk galaxies
were formed at $z\approx 4$, when the $h/R$ ratio was higher
than today. Therefore, their field patterns may have been
substantially different from those of local galaxies (Moss
\& Sokoloff 2008), and their coherence scales were still smaller
than their sizes. Our model predicts that dwarf and giant galaxies have generate regular fields of several $\mu$G strength by $z\approx 4$, while MW-type galaxies should not host these fields
before $z\approx 3$. An important implication is that polarized
radio disks and Faraday rotation are expected to exist in all
galaxies at high redshifts ($z\la3$). Even if these galaxies
cannot be resolved by the SKA beam, the integrated radio emission should be
significantly polarized for moderate and high inclinations (Stil et al. 2008). A statistical analysis of a significant sample of galaxies observed by
the SKA surveys would be able to test the validity of the predictions of this paper. The detailed field pattern can be measured with the help of a dense RM grid of polarized background sources (Stepanov et al. 2008).

Major merging events destroy the field regularity, while the total
equipartition field strength is increased over a certain period (see
Fig.~\ref{fig:mfe1} and Sect.~\ref{subsec:isfm}). Furthermore,
violent forces during the merging process can compress the turbulent
field and lead to strong polarized emission. Hence, the polarized
intensity increases, although the large-scale coherent field will be weaker than before. Faraday rotation measurements are
required to distinguish coherent from anisotropic fields.

Some fraction of galaxies will be observed in a post-merger or
starburst phase when the regular field has been temporarily destroyed and
has not yet recovered (Fig.~\ref{fig:mfe1}). A correlation between
Faraday rotation and signatures of a recent merger is expected and
should be investigated, whereas the total and polarized emission
should be higher due to enhanced turbulence and a fast outflow. Our
estimates constrain the time since the peak of the
starburst and will be useful for observing distant, starburst galaxies.

Outflows from starburst galaxies contribute to the magnetization of
the intergalactic medium (IGM) (Kronberg et al. 1999). Estimating
the importance of this effect will require reliable statistical data about starburst frequency, e.g. from radio observations with SKA.

\section{Conclusions}
\label{sec:conclusions}

Studying the evolution in magnetic fields of galaxies is important for interpreting future radio synchrotron observations with the
planned Square Kilometre Array (SKA). We have used the dynamo theory to
derive the timescales of amplification and ordering of magnetic
fields in disk and quasi-spherical galaxies. This has provided a useful tool in developing a simple evolutionary model of regular magnetic fields, coupled with models describing the formation and evolution of galaxies. In the
epoch of \emph{dark matter halo formation}, seed magnetic fields of
$\sim10^{-18}$~G strength were generated in protogalaxies by the
Biermann battery. Turbulence in the protogalactic halo generated by
thermal virialization could have driven the turbulent (small-scale) dynamo
and amplify the seed field to the equipartition level of
$\approx 20~\mu$G within a few $10^8$ yr. In the epoch of
\emph{disk formation}, the turbulent field served as a seed for the
mean-field (large-scale) dynamo developed in the disk.

\begin{itemize}

\item We defined three characteristic timescales for the evolution
of galactic magnetic fields: one for the amplification of the seed
field, a second for the amplification of the large-scale regular field, and a third
for the field ordering on the galactic scale.

\item Galaxies similar to the Milky Way formed their disk at
$z\approx10$. Regular fields of equipartition (several $\mu$G) strength and a few kpc coherence length were generated within 2~Gyr (until $z\approx3$), but field ordering up to the coherence scale of
the galaxy size took another 6~Gyr (until $z\approx0.5$).

\item Giant galaxies had already formed their disk at
$z\ga10$, allowing more efficient dynamo generation of equipartition regular
fields (with a coherence length of about 1~kpc) until
$z\approx4$. However, the age of the Universe is too young for fully coherent fields to have already developed in giant galaxies larger than about 15~kpc.

\item Dwarf galaxies formed even earlier and should have
hosted fully coherent fields at $z\approx1$.

\item Major mergers excited starbursts with enhanced
turbulence, which in turn amplified the turbulent field, whereas the regular
field was disrupted and required several Gyr to recover. Measurement of
regular fields can serve as a clock for measuring the time since the
last starburst event.

\item Starbursts due to major mergers enhance the
turbulent field strength by a factor of a few and drive a fast wind outflow, which magnetizes the intergalactic medium. Observations of the radio emission from distant starburst galaxies can provide an
estimate of the total magnetic-field strength in the IGM.

\end{itemize}

This evolutionary scenario can be tested by measurements of
polarized synchrotron emission and Faraday rotation with the SKA. We predict an anticorrelation at fixed redshift between galaxy size and the ratio between ordering scale and the galaxy size.
Weak regular fields (small Faraday rotation) in galaxies at $z\la3$,
possibly associated with strong anisotropic fields (strong polarized
emission), would be signatures of major mergers. Undisturbed dwarf
galaxies should host fully coherent fields, giving rise to strong
Faraday rotation signals.

\begin{acknowledgements}
This work is supported by the European Community Framework Programme
6, Square Kilometre Array Design Study (SKADS) and the DFG-RFBR
project under grant 08-02-92881. We acknowledge valuable discussions
with Lucio Mayer, Klaus Dolag and John Wise who provided also
simulations of the density and power spectrum of halos. We thank
Philip Kronberg for careful reading the manuscript and valuable
discussions.
\end{acknowledgements}


\begin{thebibliography}{}

\bibitem{}
Batchelor, G.~K. 1950, Proc. R. Soc. Lond., A201, 405

\bibitem{}
Baugh, C.~M., Cole, S., \& Frenk, C.~S. 1996, \mnras, 283, 1361

\bibitem{}
Beck, R. 2005, in Cosmic Magnetic Fields, eds. R. Wielebinski \& R.
Beck, Springer, Berlin, p.~43

\bibitem{}
Beck, R., Poezd, A.~D., Shukurov, A., \& Sokoloff, D.~D. 1994, A\&A,
289, 94

\bibitem{}
Beck, R., Brandenburg, A., Moss, D., Shukurov, A., \& Sokoloff, D.
1996, Ann. Rev. Astron. Astrophys., 34, 155

\bibitem{}
Beck, R., Shukurov, A., Sokoloff, D., \& Wielebinski, R. 2003, A\&A,
411, 99

\bibitem{}
Belved\'ere, G., Lanza, A., \& Sokoloff, D. 1998, Solar Physics,
183, 435

\bibitem{}
Bernet, M.~L., Miniati, F., Lilly, S.~J., Kronberg, P.~P., \&
Dessauges-Zavadsky, M. 2008, Nature, 454, 302

\bibitem[Bianchi et al.(2001)]{}
Bianchi, S., Cristiani, S., \& Kim, T.-S.\ 2001, A\&A, 376, 1

\bibitem{}
Binney, J., \& Tremaine, S. 1987, Galactic dynamics,  Princeton
University Press

\bibitem{}
Birnboim, Y., \& Dekel, A. 2003, \mnras, 345, 349

\bibitem{}
Bournaud, F., Jog, C.~J., \& Combes, F. 2007, A\&A, 476, 1179

\bibitem{}
Brook, C.~B., Kawata, D., Gibson, B.~K., \& Freeman, K.~C. 2004,
\apj, 612, 894

\bibitem{}
Chy\.zy, K.~T., Bomans, D.~J., Krause, M., et al. 2007, A\&A, 462,
933

\bibitem{}
Combes, F. 2005, in The Evolution of Starbursts, American Institute
of Physics Conference Series, 783, 43

\bibitem{}
Condon, J.~J., Anderson, M.~L., \& Helou, G. 1991, \apj, 376, 95

\bibitem[Courteau et al.(2007)]{}
Courteau, S., Dutton, A.~A., van den Bosch, F.~C., MacArthur, L.~A.,
Dekel, A., McIntosh, D.~H., \& Dale, D.~A.\ 2007, \apj, 671, 203

\bibitem[Davies \& Widrow(2000)]{}
Davies, G., \& Widrow, L.~M.\ 2000, \apj, 540, 755

\bibitem[de Avillez\& Breitschwerdt(2005)]{}
de Avillez, M.~A., \& Breitschwerdt, D.\ 2005, A\&A, 436, 585

\bibitem{}
Dekel, A., \& Birnboim, Y.\ 2006, \mnras, 368, 2

\bibitem{}
di Matteo, P., Combes, F., Melchior, A.-L., \& Semelin, B. 2007,
A\&A, 468, 61

\bibitem{}
Dib, S., Bell, E., \& Burkert, A. 2006, \apj, 638, 797

\bibitem{}
Elmegreen, B.~G., Elmegreen, D.~M., Vollbach, D.~R., Foster, E.~R.,
\& Ferguson, T.~E. 2005, \apj, 634, 101

\bibitem{}
Elmegreen, D.~M., Elmegreen, B.~G., Ravindranath, S., \& Coe, D.~A.
2007, \apj, 658, 763

\bibitem{}
Fletcher, A., Berkhuijsen, E.~M., Beck, R., \& Shukurov, A. 2004,
A\&A, 414, 53

\bibitem{}
Frick, P., Stepanov, R., \& Sokoloff, D. 2006, Phys. Rev. E, 74,
066310

\bibitem{}
Gaensler, B.~M., Beck, R., \& Feretti, L. 2004, New Astr. Rev., 48,
1003

\bibitem{}
Gaensler, B.~M., Haverkorn, M., Staveley-Smith, L., et al. 2005,
Science, 307, 1610

\bibitem{}
Genzel, R., Tacconi, L. J., Eisenhauer, F. et al. 2006, Nature, 442,
786

\bibitem{}
Governato, F., Willman, B., Mayer, L., Brooks, A., Stinson, G.,
Valenzuela, O., Wadsley, J., \& Quinn, T. 2007, \mnras, 374, 1479

\bibitem{}
Greif, T.~H., Johnson, J.~L., Klessen, R.~S., \& Bromm, V. 2008,
arXiv:0803.2237

\bibitem{}
Gressel, O., Elstner, D., Ziegler, U., \& R\"udiger, G. 2008, A\&A,
486, L35

\bibitem{}
Hanasz, M. Lesch, H., Otmianowska-Mazur, K., \& Kowal, G. 2005, in
The Magnetized Plasma in Galaxy Evolution, eds. K.~T. Chyzy,
K.~Otmianowska-Mazur, M. Soida, \& R.-J. Dettmar, Jagiellonian
University, Krakow, p.~162

\bibitem{}
Harrison, E.~R. 1970, \mnras, 147, 279

\bibitem{}
Helou, G., \& Bicay, M.~D. 1993, \apj, 415, 93

\bibitem{}
Iskakov, A.~B., Schekochihin, A.~A., Cowley, S.~C., McWilliams, J.~
C., Proctor, M.~R.~E., 2007, Physical Review Letters, vol. 98, 501,
2007

\bibitem{}
Ivison, R.~J., Smail, I., Dunlop, J.~S., et al. 2005, \mnras, 364,
1025

\bibitem{}
Julian, W.~H., \& Toomre, A. 1966, \apj, 146, 810

\bibitem[Kallivayalil et al.(2006)]{}
Kallivayalil, N., van der Marel, R.~P., \& Alcock, C.\ 2006, \apj,
652, 1213

\bibitem{}
Kauffmann, G., White, S.~D.~M., \& Guiderdoni, B. 1993, \mnras, 264,
201

\bibitem{}
Kaufmann, T., Wheeler, C., \& Bullock, J.~S. 2007, \mnras, 382, 1187

\bibitem{}
Kazantsev, A.~P. 1967, JETP, 53, 1806 [Sov. Phys. JETP, 26, 1031,
1968]

\bibitem{}
Kennicutt, R.~C., Jr., Roettiger, K.~A., Keel, W.~C., van der Hulst,
J.~M., \& Hummel, E. 1987, \aj, 93, 1011

\bibitem{}
Kitchatinov, L.~L., \& R\"udiger, G. 2004, A\&A, 424, 565

\bibitem{}
Kleeorin, N., Moss, D., Rogachevskii, I., \& Sokoloff, D. 2002,
A\&A, 387, 453

\bibitem{}
Klein, U., Wielebinski, R., \& Morsi, H.~W. 1988, A\&A, 190, 41

\bibitem{}
Kolmogorov, A.~N., Petrovsky, I.~G., \& Piskunov, N.~S. 1937, Bull.
Moscow Univ., A1, 1

\bibitem[Korpi et al.(1999)]{}
Korpi, M.~J., Brandenburg, A., Shukurov, A., Tuominen, I., \&
Nordlund, {\AA}.\ 1999, \apjl, 514, L99

\bibitem{}
Kowal, G., Otmianowska-Mazur, K., Hanasz, M. 2006, A\&A, 445, 915

\bibitem{}
Krause, M. 1990, in Galactic and Intergalactic Magnetic Fields, eds.
R. Beck, R. Wielebinski, P.~P. Kronberg, IAU Symp., 140, 187

\bibitem{}
Krause, M. 2008, arXiv:astro-ph/0806.2060

\bibitem{}
Kronberg, P.~P. 1994, Rep. Progr. Phys., 57, 325

\bibitem{}
Kronberg, P.~P., Perry, J.~J., \& Zukowski, E.~L.~H. 1992, \apj,
397, 528

\bibitem{}
Kronberg, P.~P., Lesch, H., \& Hopp, U. 1999, ApJ, 511, 56

\bibitem{}
Kronberg, P.~P., Bernet, M.~L., Miniati, F., Lilly, S.~J., Short,
M.~B., \& Higdon, D.~M. 2008, \apj, 676, 70

\bibitem{}
Kulsrud, R.~M. 1999, Ann. Rev. Astron. Astrophys., 37, 37

\bibitem{}
Landau L.~D., \& Lifshitz, E.~M. 1959, Fluid Mechanics,
Addison-Wesley, Reading, Mass

\bibitem{}
Labb{\'e}, I., Rudnick, G., Franx, M., et al. 2003, \apjl, 591, L95

\bibitem{}
Le F{\`e}vre, O., Abraham, R., Lilly, S.~J., et al. 2000, \mnras,
311, 565,

\bibitem{}
Lesch, H., \& Chiba, M. 1995, A\&A, 297, 305

\bibitem{}
Lisenfeld, U., V\"olk, H.~J., \& Xu, C. 1996, A\&A, 314, 745

\bibitem{}
Mayer, L., Governato, F., \& Kaufmann, T. 2008, arXiv:0801.3845

\bibitem{}
Medvedev, M.~V., Silva, L.~O., Fiore, M., Fonseca, R.~A., \& Mori,
W.~B. 2004, J. Korean Astron. Soc., 37, 533

\bibitem{}
Meneguzzi, M., Frisch, U., Pouquet, A. 1981, Phys. Rev. Lett., 47,
1060

\bibitem{}
Mishustin, I.~N., \& Ruzmaikin, A.~A. 1971, Sov. Phys. JETP,  61,
441

\bibitem{}
Moss, D., Snodin, A.~P., Englmaier, P., Shukurov, A., Beck, R., \&
Sokoloff, D. 2007, A\&A, 465, 157

\bibitem{}
Moss, D., \& Sokoloff, D.\ 2008, A\&A, 487, 197

\bibitem{}
Moss, D., Shukurov, A. \& Sokoloff, D. 2008, Geophys. Astrophys.
Fluid Dyn., 89, 285

\bibitem{}
Niklas, S., \& Beck, R. 1997, A\&A, 320, 54

\bibitem{}
Omukai, K., \& Palla, F. 2003, \apj, 589, 677

\bibitem[Piatek et al.(2008)]{}
Piatek, S., Pryor, C., \& Olszewski, E.~W.\ 2008, \aj, 135, 1024

\bibitem{}
Press, W.~H., \& Schechter, P. 1974, ApJ, 187, 425

\bibitem[Pudritz \& Silk(1989)]{}
Pudritz, R.~E., \& Silk, J.\ 1989, ApJ, 342, 650

\bibitem{}
Rees, M.~J. 2006, Astron. Nachr., 327, 395

\bibitem[Rees\& Ostriker(1977)]{}
Rees, M.~J., \& Ostriker, J.~P.\ 1977, MNRAS, 179, 541

\bibitem[Reshetnikov et al.(2003)]{}
Reshetnikov, V.~P., Dettmar, R.-J., \& Combes, F. 2003, A\&A, 399,
879

\bibitem{}
Reuter, H.-P., Klein, U., Lesch, H., Wielebinski, R., \& Kronberg,
P.~P. 1994, A\&A, 282, 724

\bibitem{}
Roussel, H., Helou, G., Beck, R., et al. 2003, \apj, 593, 733

\bibitem{}
R\"udiger, G., \& Hollerbach, R. 2004, The Magnetized Universe.
Geophysical and Astrophysical Dynamo Theory, Wiley-VCH, ESO

\bibitem{}
Ruzmaikin, A., Shukurov, A., \& Sokoloff, D. 1988, Magnetic Fields
of Galaxies, Kluwer, Dordrecht

\bibitem{}
Ryan, R.~E., Jr., Cohen, S.~H., Windhorst, R.~A., \& Silk, J.\ 2008,
\apj, 678, 751

\bibitem{}
Schlickeiser, R., \& Shukla, P.~K. 2003, \apj, 599, L57


\bibitem{}
Semikoz, V.~B., \& Sokoloff, D.~D. 2005, Int. J. Modern Phys. D, 14,
1839

\bibitem{}
Seymour, N., Dwelly, T., Moss, D., et al. 2008, \mnras, 386, 1695

\bibitem{}
Shukurov, A. 2004, in Mathematical aspects of Natural Dynamos,
arXiv:astro-ph/0411739

\bibitem{}
Shukurov, A. 2005, in Cosmic Magnetic Fields, eds. R. Wielebinski \&
R. Beck, Springer, Berlin, p.~113

\bibitem{}
Sokoloff, D. 1995, Magnetohydrodynamics,  31, 43

\bibitem{}
Sokoloff, D. 2002,  Astron. Rep., 96, 871

\bibitem{}
Sokoloff D.~D., Nefedov S.~N., Ermash A.~A., \& Lamzin, S.~A. 2008,
arXiv:0806.0746

\bibitem{}
Sokoloff, D., \& Shukurov, A. 1990, Nature, 347, 51

\bibitem{}
Springel, V., \& Hernquist, L. 2005, \apjl, 622, L9

\bibitem{}
Stepanov, R., Arshakian, T.~G., Beck, R., Frick, P., \& Krause, M.
2008, A\&A, 480, 45

\bibitem{}
Stil, J.~M., Krause, M., Beck, R., \& Taylor, A.~R.\ 2008, arXiv:0810.2303

\bibitem{}
Stringer, M.~J., \& Benson, A.~J. 2007, \mnras, 382, 641

\bibitem{}
Subramanian, K. 1998, \mnras, 294, 718

\bibitem{}
Takahashi, K., Ichiki, K., Ohno, H., Hanayama, H., \& Sugiyama, N.
2006, Astron. Nachr., 327, 410

\bibitem{}
Takizawa, M. 2005, \apj, 629, 791

\bibitem[Tegmark et al.(1997)]{}
Tegmark, M., Silk, J., Rees, M.~J., Blanchard, A., Abel, T., \&
Palla, F. 1997, \apj, 474, 1

\bibitem[Tobias \& Cattaneo, F. (2008)]{}
Tobias, S.~M., \& Cattaneo, F.\ 2008, Phys. Rev. Lett, 101, 125003

\bibitem[Trujillo et al.(2006)]{}
Trujillo, I., et al. 2006, \apj, 650, 18

\bibitem{}
van der Hulst, J.~M., Sadler, E.~M., Jackson, C.~A., et al. 2004,
New Astr. Rev., 48, 1221

\bibitem{}
Villalobos, {\'A}., \& Helmi, A. 2008, arXiv:0803.2323

\bibitem{}
Vollmer, B., Soida, M., Beck, R., et al. 2007, A\&A, 464, L37

\bibitem[Wang \& Abel(2007)]{}
Wang, P., \& Abel, T.\ 2007, arXiv:0712.0872

\bibitem{}
Weibel, E.~S. 1959, Phys. Rev. Lett. 2, 83

\bibitem{}
Wielebinski, R. 2006, Astron. Nachr. 327, 510

\bibitem{}
Wezgowiec, M., Urbanik, M., Vollmer, B., et al. 2007, A\&A, 471, 93

\bibitem{}
Wise, J.~H., \& Abel, T. 2007, \apj, 665, 899

\bibitem{}
Wise, J.~H., Matthew, J., Turk, M.~J., \& Abel, T. 2007,
arXiv:0710.1678

\bibitem{}
Wolfe, A.~M., Jorgenson, R.~A., Robishaw, T., Heiles, C., \&
Prochaska, J.~X. 2008, Nature, 455, 638

\bibitem[Xu et al.(2008)]{}
Xu, H., O'Shea, B.~W., Collins, D.~C., Norman, M.~L., Li, H., \& Li,
S.\ 2008, arXiv:0807.2647

\bibitem{}
Zeldovich, Ya.~B., Ruzmaikin, A.~A., \& Sokoloff, D.~D. 1990, The
Almighty Chance, Singapore, World Sci.


%
%
%
%
%
%
%
%
\end{thebibliography}
\end{document}